# Room temperature quantum metric effect in TbMn$_6$Sn$_6$

Weiyao Zhao[1, 9], Kaijian Xing[2, 9], Yufei Zhao[3, 9], Lei Chen[4], Min Hong[4], Yuefeng Yin[1], Yang Liu[5], Khoa Dang Le[6], Jacob Gayles[6], Fang Tang[7], Yong Fang[7], Binghai Yan[3,8], and Julie Karel[1*]

[1] *Department of Materials Science & Engineering, & ARC Centre of Excellence in Future Low-Energy Electronics Technologies, Monash University, Clayton VIC 3800, Australia*

[2] *School of Physics & Astronomy, Monash University, Clayton VIC 3800, Australia*

[3] *Department of Condensed Matter Physics, Weizmann Institute of Science, Rehovot 7610001, Israel*

[4] *Centre for Future Materials, University of Southern Queensland, Springfield QLD 4300, Australia*

[5] *Monash Centre for Electron Microscopy, Monash University, Clayton VIC 3800, Australia*

[6] *Department of Physics, University of South Florida, Tampa, FL 33620, USA*

[7] *Jiangsu Laboratory of Advanced Functional Materials, Changshu Institute of Technology, Changshu 215500, China*

[8] *Department of Physics, The Pennsylvania State University, University Park, Pennsylvania 16802, USA*

[9] *The authors contribute equally to this work.*

Quantum geometry, including Berry curvature and the quantum metric, of the electronic Bloch bands has been studied via nonlinear responses in topological materials. Naturally, these material systems with intrinsic strong nonlinear responses also form the key component in nonlinear electronic devices. However, the previous reported quantum geometry effects are mainly observed at cryogenic temperatures, hindering their application in practical devices. Here we report the tuneable strong room-temperature second-harmonic transport response in a quantum magnet, TbMn$_6$Sn$_6$, which is governed by the quantum metric and can be tuned with applied magnetic fields. We show that around room temperature, which is close to the spontaneous spin-reorientation transition, the magnetic configurations, and therefore the related symmetry breaking phases, are easily controlled via magnetic fields. Our results also show that manipulation of the symmetries of the magnetic structure presents an effective route to tuneable quantum-geometry-based devices.

* julie.karel@monash.edu

Nonlinear responses, e.g., high-harmonic electric/optical generation, in quantum materials have been employed to trigger promising applications such as rectification and terahertz detection(*1-4*). Nonlinear transport also provides valuable insight into the intricate quantum geometry in condensed matter physics(*5-7*). In momentum space, the real and imaginary part of the quantum geometry defines the metric and curvature of Bloch wavefunctions, respectively. For example, in non-magnetic WTe$_2$ flakes, the nonlinear anomalous Hall effect is caused by the Berry curvature dipole ($D_{BC}$) (*8*); in antiferromagnetic MnBi$_2$Te$_4$ thin flakes, the quantum metric dipole ($D_{metric}$) gives rise to both transverse and longitudinal nonreciprocal responses below 24 K (*9, 10*). The electric magneto-chiral anisotropy (eMChA) in chiral crystals (*11*), described by $R(I,B) = R_0(1 + \mu^2 B^2 + \gamma^{\pm} B \cdot I)$, also has a $D_{metric}$ contribution(*12*). The nonlinear response is commonly characterized by the second-harmonic voltage $V^{2\omega}$ (Fig. 1A). Because $D_{BC}$ is even under time-reversal symmetry ($T$), the resultant $V^{2\omega}$ remains unchanged when the magnetic order reverses. In contract, the $T$-odd $D_{metric}$ -induced $V^{2\omega}$ changes sign if magnetic order flips. Figure 1A shows a generic nonlinear $V^{2\omega}$ dependence on $I^{2\omega}$ when $D_{BC}$ and $D_{metric}$ coexist.

The nonlinear transport is sensitive to inversion ($P$) -breaking and $T$-breaking and provides an excellent tool to probe hidden symmetry-breaking phases. One prime example is the charge-ordered kagome metal CsV$_3$Sb$_5$, which is generally viewed with a centrosymmetric lattice distortion ($T_{CDW}$ = 94 K) but exhibits chiral and $T$-breaking quasi-particle interference(*13-15*). The eMChA was observed below 35 K(*13*), indicating the possible presence of orbital loop currents, which spontaneously reduces symmetry. In MnBi$_2$Te$_4$ and CsV$_3$Sb$_5$, while the nonlinear response can be manipulated by an external magnetic field, it only survives at low temperatures (*9, 10*), which is of limited use in practical device applications.

In this work, we report unexpected second order nonlinear electronic transport at room temperature in the centrosymmetric magnet TbMn$_6$Sn$_6$, which holds rich magnetic interactions. The response has a complicated temperature dependence driven by the spin-reorientation transition around room temperature (e.g., 270 – 330 K) (*16-19*). More importantly, we show that the field-dependent $D_{metric}$ is tuneable via controlling the magnetic configuration and is accompanied by a relatively-large field-independent response. Our results pave the way for nonlinear device design based on quantum magnets operating at room temperature.

**The magnetic phases in TbMn₆Sn₆** Modifying the crystal symmetry in a quantum material is one of the most important routes to manifesting sizeable $D_{BC}$ and $D_{metric}$ -induced nonlinear physics, such as that observed in artificially corrugated bilayer graphene.(*20-23*) An alternative approach in nonlinear responses is to introduce magnetic order to break symmetries. The kagome magnet system, such as the $R$Mn₆Sn₆ family, where $R$ is a rare earth element, is a promising candidate for locating this magnetism-induced symmetry-breaking behaviour. $R$Mn₆Sn₆ compounds crystallize in a HfFe₆Ge₆-type structure (centrosymmetric space group *P*6/*mmm*), which possesses Mn kagome layers stacked along the *c* axis. In each unit cell, the $R$ atoms occupy the honeycomb centres between two Mn layers. Except for the case of a nonmagnetic $R$ (Y and Lu) atom, the competing magnetic interactions, namely direct Mn-Mn exchange, indirect 4*f*-3*d* type $R$-Mn exchange and Ruderman-Kittel-Kasuya-Yosida (RKKY) type $R$-$R$ interactions, contribute to rather complicated magnetic structures, offering a rich field for nonlinear physics. Among these magnetic interactions, intralayer Mn-Mn interactions are the strongest, and therefore ferromagnetic coupling within the kagome planes dominates. The hybridization between Mn 3*d* and $R$ 5*d* orbitals always favours antiparallel alignment, thus yielding a negative exchange coupling between Mn and $R$(*24*). In $R$Mn₆Sn₆, various magnetic structures have been observed, such as an easy-*ab*-plane anisotropy in GdMn₆Sn₆, an easy-*c*-axis anisotropy in TbMn₆Sn₆ and conical order in DyMn₆Sn₆ and HoMn₆Sn₆(*25*). Particularly, in TbMn₆Sn₆, the ferrimagnetic ordering forms below $T_C$ ~ 420 K, with both Tb and Mn magnetic moments aligned parallel/antiparallel to the *a*-axis, and, a spin reorientation(*18, 19*) occurs at $T_{SR}$ ~ 315 K, where the moments rotate into the *c*-axis with a slight canting angle, as shown in Fig. 1B, (more SR transition sketch and related magnetization measurements are available in Figs. S1 & S2). The canting effect is supported by magnetization measurements shown in Fig. S2. The magnetization does not saturate at high magnetic field, and the in-plane and out-of-plane magnetic moments at $B$ = 5 T at 300K are different, e.g., 5.25 $\mu_B$/*f. u.* and 5.19 $\mu_B$/*f. u.*, respectively. These results confirm the robust canted ferrimagnetic order and in-plane antiferromagnetic components at room temperature. The magnetic ordering in the kagome lattice contributes to large Berry curvature in the material, which leads to a giant anomalous Nernst effect(*26, 27*) and anomalous Hall effect(*25, 28*) at around room temperature (*16, 19, 25-27, 29-37*). On top of this, inelastic neutron scattering (INS) results(*16, 17, 19*) indicate that the Mn in-plane spin configuration of the kagome lattice has magnetic excitation features of ferromagnetism, flat-band antiferromagnetism, and chiral antiferromagnetism (Fig. 1B). Importantly for this work, the flat-band type antiferromagnetic spin structure in the Mn kagome lattice, with an inversion-symmetric pair of Mn moments canting along opposite directions, result in breaking *P* and *T*, satisfying conditions to observe the quantum metric physics.

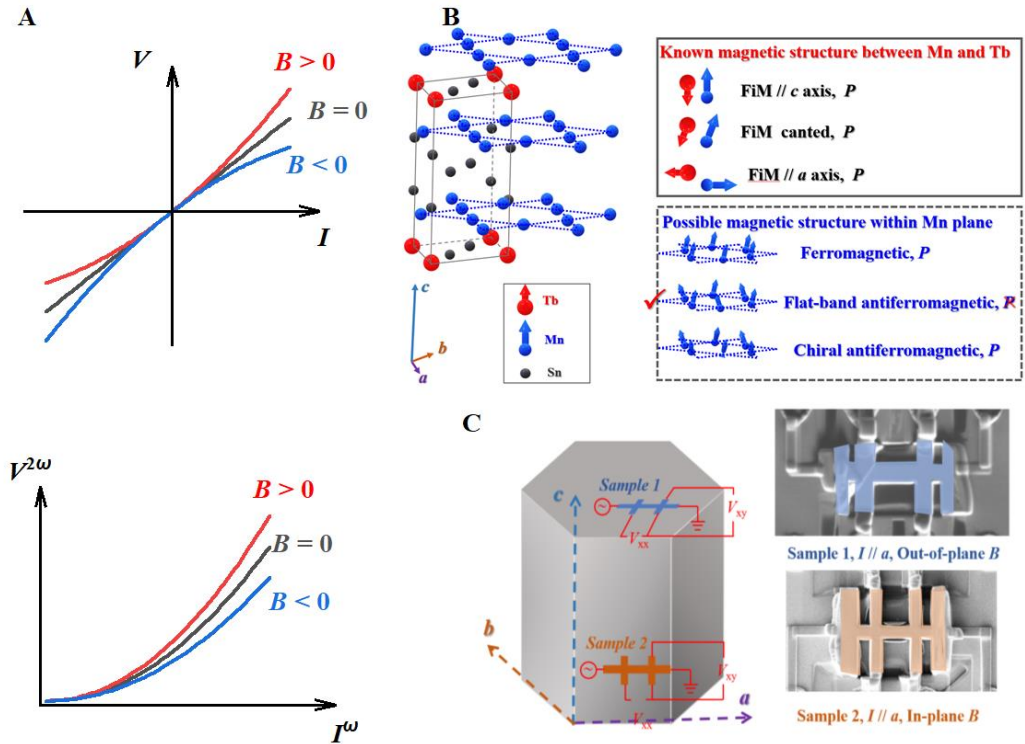

Fig. 1 (A) The nonlinear magnetoelectric responses, in (top) first order, e.g., a chiral conductor showing electric magneto-chiral anisotropy in applied magnetic fields, and (bottom) second order, e.g., the parabolic second-harmonic *I-V* curves show magnetic field dependence due to the quantum metric and Berry curvature dipole contributions. The second-order conductivity induced by $D_{BC}$ and $D_{metric}$ have even and odd time-reversal symmetry (*T*) dependence, respectively, therefore will contribute to different parabolic *I-V* responses. (B) In the quantum magnet TbMn$_6$Sn$_6$, alternative stacking of kagome lattice Mn layers and honeycomb lattice Tb layers form ferrimagnetic ordering. The temperature dependent ferrimagnetic orders are sketched in solid-line box between two sublattices, which is the well-known magnetic configuration. Note that, the middle configuration in the solid box shows room-temperature magnetic moments' orientation for Mn and Tb, in which the tilting angle is slightly exaggerated [~ 13° at 300 K(*29*)] for a clearer demonstration of the tilting effect. The "hidden" local spin texture of the Mn-kagome plane are sketched in the dash-line box, which are the possible local structures as verified by INS(*17*), in which the so-called "flat-band antiferromagnetic" in-plane components break the inversion symmetry in the Mn-kagome plane. Thus, the quantum metric dipole is nonzero due to the broken *T* and inversion (*P*) symmetries. (C) The demonstration of focused-ion beam (FIB) fabrication devices in TbMn$_6$Sn$_6$ crystal. Both samples have current flowing along the *a* axis, while the orientation of applied fields are along the out-of-plane (*c* axis) direction and within the basal plane ([$\bar{1}\bar{2}0$] direction), respectively.

**Electronic transport properties of TbMn$_6$Sn$_6$** The anisotropic magnetotransport properties of TbMn$_6$Sn$_6$ with applied field along the *c* axis or the basal plane ([$\bar{1}\bar{2}0$] direction) for Sample 1 and Sample 2 (see Fig. 1C for sample orientation), respectively, have been studied (Figs. S3 & S4), indicating the excellent crystal quality of the studied devices. Figs. 2A & B demonstrates the first-order transport results at 300 K. The negative magnetoresistance (MR) in Sample 1 is sharper at low fields, where the moments are not fully aligned, and shows linear dependency with fields above the alignment, indicating spin-disorder-scattering and electron-magnon scattering, respectively(*38*). With applied in plane fields, a field-induced SR occurs, and shows a U-shape positive MR in the SR region, similar to the reports in the Fe$_3$Sn$_2$ kagome magnet(*39*). A giant anomalous Hall effect arising from the large Berry curvature in the band structure (*27, 28, 37*) is also observed in Samples 1 & 2, as shown in Fig. 2B. The anomalous Hall conductivity $\sigma_{xy}^A$ $(= \frac{\rho_{xy}^A}{\rho_{xy}^2 + \rho_{xx}^2})$ is ~ 210 S·cm$^{-1}$ in sample 1, which converts to ~ 0.122 $e^2/h$ (quantum conductance) per kagome plane, comparable with top-quality TbMn$_6$Sn$_6$ crystals in literature. In Sample 2, the similar result (0.13 $e^2/h$) is also observed at room temperature, implying the Berry curvature may also contribute to an out-of-kagome plane giant anomalous Hall conductivity. The observed similar linear transport behavior in Sample 1 and Sample 2 also suggests that the electronic properties in TbMn$_6$Sn$_6$ are not affected by the FIB fabrication process. Similar behavior can be found in the 290 – 320 K region, as demonstrated in Fig. S3; the calculated anomalous Hall conductivity values are summarized in Fig. 2G. The anomalous conductivity is roughly a constant, and independent with the longitudinal conductivity, also suggesting the intrinsic origin from the Berry curvature.

**Nonlinear transport at room temperature** A frequency doubling effect was observed in the FIB-fabricated Hall bar samples by the application of an alternating current (a.c.), as shown in Fig. 2C-F for Sample 1, and Fig. S5 for Sample 2. Figs. 2 C&D show both longitudinal and transverse second harmonic responses $V_{xx}^{2\omega}, V_{xy}^{2\omega}$. Note that, the second harmonic transport behavior contains two parts: 1) a magnetic-field-independent offset, and 2) the magnetic-field-dependent $\Delta V^{2\omega}$ which can be obtained via subtracting a constant offset value at $B = 0$ T. Based on these different contributions, the nonlinear response signals have different *T*-dependence, e.g., the Berry curvature dipole is a *T*-even term while the Drude and quantum metric are *T*-odd terms.(*40*) Therefore, the $D_{\text{metric}}$-induced term can be, in principle, obtained via analysing the magnetic field dependence of the second harmonic transport properties. After subtracting the *T*-even term of each measured curve in Sample 1, $\Delta V_{xx}^{2\omega}, \Delta V_{xy}^{2\omega}$ are obtained and plotted in conductance per Mn-kagome plane. (H) The longitudinal and (I) transverse second-order responses at different magnetic fields are plotted against the applied a.c. density along the *a* axis.

Figs. 2 E&F. $\Delta V_{xx}^{2\omega}$ and $\Delta V_{xy}^{2\omega}$ show nonlinear behavior with magnetic field, which are odd functions and appear to saturate at sufficiently large fields. Hence, it is highly possible that both $D_{BC}$ and $D_{metric}$ contribute to the nonlinear transport behavior. Further, we summarize the longitudinal and transverse nonlinear response at $B = 0, \pm 1$ T values as functions of the a.c. density (I-V) in Fig. 2 H&I, for Sample 1. The parabolic fitting curves show the current dependency of the second harmonic transport signals and clearly indicate the signature of a second-order electronic transport response. The splitting at different fields shows the tuneability in the ($D_{BC}$ and $D_{metric}$) co-contribution model.

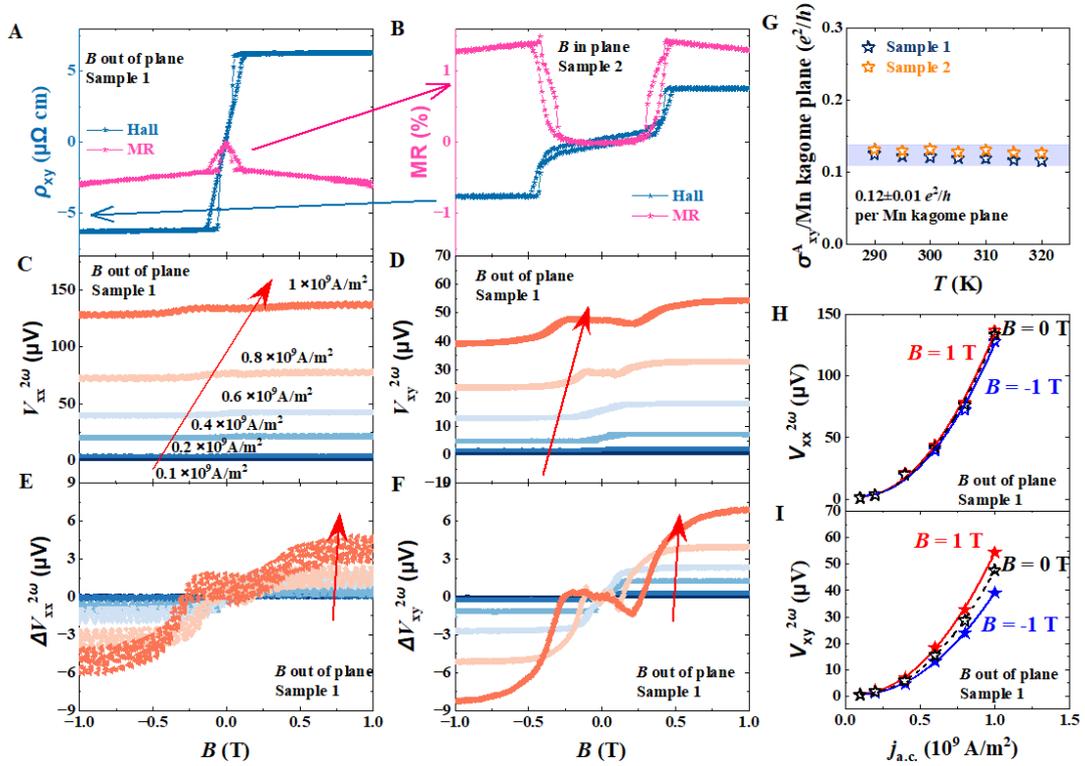

Fig. 2 The anisotropic transport properties of TbMn$_6$Sn$_6$ at 300 K. (A, B) The first order magnetotransport properties of Sample 1 and Sample 2 in -1 – 1 T range, in which the Hall effect of both samples are plotted in blue curve and use the blue *y* axis in Panel (A), the MR curves are plotted in purple and share the purple *y* axis in Panel (B). (C, D) The nontrivial longitudinal and transverse second harmonic response in Sample 1 with different applied a.c. densities. (E, F) The magnetic-field dependent components in the nonlinear response shown in Panel C&D, respectively. Note that, Panel (C-F) share the same color code for current density level, which is labelled in Panel (c) and indicated by the red arrows along increasing current density directions in each Panel. (G) The Berry curvature induced anomalous Hall conductivity in 290 – 320 K region, which is converted to quantum

In the co-contribution model of the $D_{BC}$ and $D_{metric}$, symmetry breaking is present and is induced by the in-kagome-plane antiferromagnetic alignment. As shown in Fig. 1B, with flat-band type antiferromagnetic, *P* and *T* symmetries are both broken, and with chiral antiferromagnetism, the *T* and $C_3$ symmetries are broken. A nonzero quantum metric dipole $D_{metric}$ may therefore give rise to the odd-with-magnetic-field nonlinear responses, on top of the magnetic-field-even components contributed by $D_{BC}$. In contrast with sample 1, the responses in Sample 2 are substantially weaker and exhibit a different field dependence (Fig. S5), implying these nonlinear responses originate from the magnetic configurations. Before moving on to further discussion, extrinsic effects should be excluded. First, we employed a scaling analysis (*6, 9, 10*) to reveal that, the second harmonic Hall conductivity (in the high-field plateau area of Fig. 2F) is nearly independent of relaxation time $\tau$ in the Drude model, excluding non-linear Drude or impurity scattering contributions to the signal (See details in Supplementary Information Fig. S11). Moreover, the current density applied in this study is also smaller than other nonlinear transport research in FIB-based devices(*13, 41-43*); considering the comparable resistivity in these metallic materials, the Joule heating effect contribution is negligible. Then to understand these nonlinear transport behaviors near the spin reorientation region, the temperature, magnetic field and a.c. density dependent measurements are conducted in both samples.

**Spin reorientation, chirality and Skyrmions** The SR is triggered by an increase in the Tb single ion anisotropy energy with cooling(*16, 19*), which favours the *c* axis and consequently forces the Mn moments to rotate in the antiparallel (-*c* in this case) direction. Since the spontaneous $T_{SR}$ is near room temperature, an applied magnetic field will also easily alter the spin configuration in this temperature range. Moreover, in this region, the energies of different magnetic configurations are comparable, thus offering the potential for skyrmion formation, which also contributes to the nonlinear transport responses. Although the nonlinear transport is insignificant when the magnetic field is applied in the basal plane (Sample 2) compared with *B* along the *c*-axis, peaks in $\Delta V_{xy}^{2\omega}$ with ~ 40% intensity of Sample 1 maximum are observed around the field-induced SR region in sample 2 (Fig. S5). To directly compare, the field-dependent second harmonic Hall effect components (see Supplementary Figs. S6 & S7 for details) are plotted into a four-dimensional color map in Fig. 3A. Low field features (indicated with a purple arrow), which are approximately the same intensity appear in both sample 1 and

2 (e.g. *B* applied out of basal plane and in basal plane, respectively). These features appear at different temperatures depending on the sample and are consistent with a field-driven spin reorientation.

To further study the nonlinear response from the spin reorientation, Fig. 3B shows a plot of $V_{SR}^{2\omega}$, which is the SR-related second harmonic Hall component (see SI for $V_{SR}^{2\omega}$ calculation and its temperature, magnetic field and a.c. density dependency). As shown in Fig. 3B, the "peak" values (~ 4.5 µV for Sample 1 and ~ 3.5 µV for Sample 2) at 305 K for both samples are comparable, indicating a similar underlying 3D-like mechanism. In Fig. 3A, a naïve model demonstrates the nominal spin configuration. Moreover, from the phase diagram shown in Fig. 3A, the quantum metric effect and skyrmion related eMChA can be isolated below 300 K. Namely with the field applied along the *c* axis and along the *a* axis, only one effect is detected in each case (quantum metric and skyrmion eMChA, respectively).

To interpret these results, it is useful to consider the rich magnetism presented in TbMn$_6$Sn$_6$. Recent neutron scattering experiments (*17, 19, 44*) suggest that on top of the long-range ferromagnetic ordering in the Mn kagome layers, antiferromagnetic chiral magnons exist. Our magnetic study (Fig. S2) also suggests the existence of in-kagome-plane Mn moment components in the studied region, which is necessary for the presence of in-plane antiferromagnetism. Moreover, Lorentz transmission electron microscope experiments (*33, 45*) have confirmed the existence of skyrmions in TbMn$_6$Sn$_6$ at room temperature, which can be manipulated by the physical environments, e.g., temperature, magnetic fields and focused electron beam. The observed peaks in Fig. 3A are in comparable temperature and field range, which further suggests its relationship with topological spin textures. Skyrmions can be also driven by applied electric current, e.g., in the topological magnet MnSi (*46*), which leads to the deformation of skyrmions and therefore, the real-space Berry phase. Thus, the a.c. density induced peak shifts in the Fig. 3B and Fig. S6&7 can be understood using this framework. The current-induced skyrmions dynamics shows clear "turn-on" effect, e.g., with 0.2×10$^9$ A/m$^2$ applied current, there are no obvious skyrmion peaks; while in the same condition, the $D_{\text{metric}}$ induced plateaus are robust. Further, the antiferromagnetic chiral magnons contain in-plane and out-of-plane components (*19*), which support 3D skyrmion dynamics, and thus contribute to the 3D-like behaviour in Fig. 3B. Therefore, TbMn$_6$Sn$_6$ is a chiral magnet near the spin-reorientation region which could give rise to a strong eMChA effect in the second harmonic Hall response.

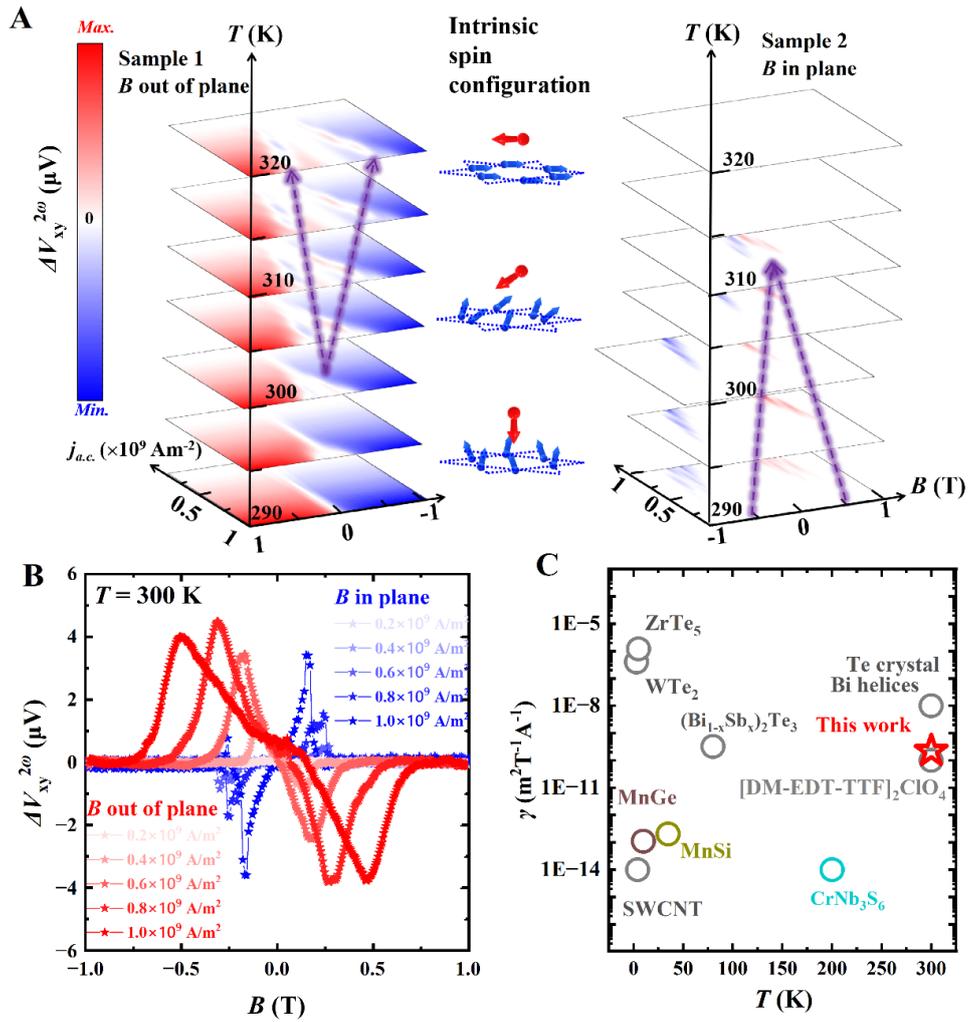

Fig. 3 Second harmonic transport in TbMn$_6$Sn$_6$. (A) The magnetic field dependent components of second harmonic Hall effect ($\Delta V_{xy}^{2\omega}$) in Sample 1 and Sample 2 are summarized in 4D color map, in which the temperature-, magnetic field-, a.c. density- axes, as well as the color code are all in the same scale for both samples. The purple arrows indicate the field induced spin reorientation transition. The schematic in Panel A, indicating the spontaneous spin orientation transition, shows the nominal spin configurations at 320, 310 and 290 K (top - down), in which the red and blue arrows indicate Tb and Mn moments, respectively. Note that, large external fields will induce the reorientation transitions, and force the moments mostly to align along the applied fields. (B) The second harmonic Hall voltage related to the chiral magnetic structure. (C) The electric magnetochiral anisotropy coefficients from some popular quantum materials(*11, 43, 47-53*) are summarized at different temperatures, in which the room-temperature maximum value observed in TbMn$_6$Sn$_6$ is highlighted. Note that, the non-tuneable compounds are plotted in grey circles, and the materials that show tuneable nonlinear transport behavior are plotted using different colors.

**Discussion and Conclusion** Based on these results, there are clearly two distinct mechanisms that contribute to the large field-dependent nonlinear response near room temperature. The first is the eMChA effect arising from the field-induced spin reorientation of the chiral magnetic structure around the spin reorientation temperature. Due to the three-dimensional nature of the skyrmions (real-space topology) in this material, the effect is observed with both out-of- and in-kagome plane applied fields. The second contribution is from the electronic band structure ($D_{metric}$ and $D_{BC}$), which arises only when $B // c$ (Fig. S9). After ruling out the current heating contribution, Drude contributions, the intrinsic anisotropic nonlinear transport result can be understood by considering the symmetry in the material.

The quantum metric can arise in a material with broken $P$ and broken $T$ symmetry. The symmetry breaking depends on in-plane antiferromagnetism, and is supported by a slight canting of the moments (~13°). (*29*) Therefore, this broken $P$ and broken $T$ symmetry is only present when $B // c$-axis. When $B //$ basal plane, the parallel alignment of the Mn moments in the kagome lattice will no longer break $P$ symmetry, and thus only skyrmion-induced components are observed in Sample 2. In this scenario, the nonlinear response can be easily controlled via changing temperatures or applied fields in room temperature region. The magnetic structure induced symmetry breaking phase is an important starting point to realize nonlinear electronic properties in quantum materials. A good example is intrinsic magnetic topological insulator $MnBi_2Te_4$, in which Mn moments are out-of-plane ferromagnetically coupled in each layer, and antiferromagnetically between adjacent layers below Néel temperature $T_N$. Below $T_N$, odd-layer $MnBi_2Te_4$ breaks both $P$ and $T$ symmetries, however preserves $PT$ symmetry, offering suitable material system to study quantum metric physics (*6*). Experimentally(*10*), $D_{metric}$ induced nonlinear transport has been observed below $T_N$. While another group shows additional symmetry manipulation via van de Waals engineering boosts the $D_{metric}$ contribution to nonlinear responses. $MnBi_2Te_4$ orders at cryogenic temperatures, and has high defect density level that may reduce the efficiency in symmetry manipulation. $TbMn_6Sn_6$ has above-room-temperature magnetism, and high crystalline quality, thus it can be an ideal platform for quantum metric physics. Recently, the studies of quantum metric induced physics at room temperature in $WTe_2$(*54*), $Mn_3Sn$(*55*) and $Cs_2Ni_3S_4$(*56*) are attracting significant attention, indicating a promising stage for implementing quantum geometry in electronic devices.

A substantial room temperature nonlinear response is a key requirement in designing highly-efficient nonlinear devices. Recently, a large $D_{BC}$ has been demonstrated using second harmonic transport at room temperature in topological materials such as the Weyl semimetal $TaIrTe_4$(*57*), the massive Dirac semimetal $BaMnSb_2$(*43*), and Bi thin films(*58*). In addition to room temperature operation, a key requirement for emerging devices is tuneability, which has been reported in a symmetry breaking phase in the Kagome CsV$_3$Sb$_5$ compound, although only at cryogenic temperatures (*13*). In Fig. 3C, the coefficient of eMChA, $\gamma = 4V^{2\omega}/(V^{\omega}Bj_{a.c.})$ in some popular materials are summarized, together with our work on TbMn$_6$Sn$_6$. Among all of the materials, TbMn$_6$Sn$_6$ is the only one which demonstrates giant, tuneable second harmonic transport behavior at room-temperature.

We have shown that in TbMn$_6$Sn$_6$, the Berry curvature dipole, quantum metric dipole and skyrmion induced second harmonic transport behavior coexist at room temperature, and can be tuned via control of the magnetic configurations. The giant and tuneable effect is confirmed in multiple samples, thus promoting a room temperature controllable response of the nonlinear physics for device design. The rich physics of kagome materials may prove a fertile playground to identify other quantum metric driven non-linear responses, which could open a new direction in practical nonlinear device design for next-generation electronics and spintronics.

**Data availability** All data supporting the findings of this study are available within the article and the Supplementary Information file, or available from the corresponding authors upon request. Source data are provided with this paper.

**Acknowledgements** This work is supported by ARC Centre of Excellence in Future Low-Energy Electronics Technologies No. CE170100039, Australian Research Council Discovery Project DP200102477, DP220103783, the National Natural Science Foundation of China No. 12174039. This work was performed in part at the Melbourne Centre for Nanofabrication (MCN) in the Victorian Node of the Australian National Fabrication Facility (ANFF). The FIB fabrications are conducted in Monash Centre for Electron Microscope (MCEM), Monash University, a Microscopy Australia (ROR: 042mm0k03) facility supported by NCRIS. The FIB equipment is funded by Australian Research Council grants (LE200100132).

**Author contributions** W.Z., K.X. and Y.Z. contributed equally to this work. W.Z. and J.K. conceived the idea. Y.F., F.T. prepared the crystals. W.Z., K.X. and Y.L. prepared the device for measurements. W.Z., Y.Z. analysed the data. W.Z., Y.Z., B.Y. and J.K. wrote the manuscript. All authors contributed to the discussion.

**Competing interests** The authors declare no competing interests.